\documentclass[prd,preprint,superscriptaddress,showpacs,byrevtex]{revtex4}
\usepackage{bm}
\def\fsl#1{\setbox0=\hbox{$#1$}           
   \dimen0=\wd0                                 
   \setbox1=\hbox{/} \dimen1=\wd1               
   \ifdim\dimen0>\dimen1                        
      \rlap{\hbox to \dimen0{\hfil/\hfil}}      
      #1                                        
   \else                                        
      \rlap{\hbox to \dimen1{\hfil$#1$\hfil}}   
      /                                         
   \fi}                                         %
\newcommand{\be}{\begin{equation}}
\newcommand{\ee}{\end{equation}}
\newcommand{\bea}{\begin{eqnarray}}
\newcommand{\eea}{\end{eqnarray}}
\newcommand{\beq}{\begin{equation}}
\newcommand{\eeq}{\end{equation}}
\newcommand{\beqs}{\begin{eqnarray}}
\newcommand{\eeqs}{\end{eqnarray}}

\newcommand{\pslash}{p\hspace{-0.067in}\slash}

\begin{document}
\title{ Operator Product Expansion in QCD Is Not Consistent With Quantum Field Theory For Gluon Distribution Function }
\author{Gouranga C Nayak }\thanks{G. C. Nayak was affiliated with C. N. Yang Institute for Theoretical Physics in 2004-2007.}
\affiliation{ C. N. Yang Institute for Theoretical Physics, Stony Brook University, Stony Brook NY, 11794-3840 USA}
\date{\today}
\begin{abstract}
Since the operator product expansion (OPE) is applicable at short distance the OPE in QCD does not solve the long distance confinement problem involving hadron in QCD where the non-perturbative QCD is applicable. In this paper we show that the gauge invariant definition of the non-perturbative gluon distribution function inside the hadron consistent with the operator product expansion (OPE) in QCD at high energy colliders is not consistent with the gauge invariant definition of the non-perturbative gluon distribution function in the quantum field theory.
\end{abstract}
\maketitle
\pagestyle{plain}
\pagenumbering{arabic}
\section{Introduction}

Quantum electrodynamics (QED) is the fundamental theory of the nature describing the interaction between electrons and photons. Similarly the quantum chromodynamics (QCD) is the fundamental theory of the nature describing the interaction between quarks and gluons. In renormalized QED the coupling decreases at long distance and in renormalized QCD \cite{ptv} the coupling decreases at short distance due to asymptotic freedom in QCD \cite{pas}.

Since the QCD coupling decreases at short distance the partonic scattering cross section at short distance is calculated by using the perturbative quantum chromodynamics (pQCD). However, since the QCD coupling increases at long distance the pQCD can not be applied to study hadron formation from partons because the hadron formation from partons involves long distance physics in QCD. Hence in order to study formation of hadron from partons the non-perturbative QCD is required at long distance where confinement happens in QCD.

Since the non-perturbative QCD is not solved yet, most of the non-perturbative quantities in QCD (such as the parton
distribution function (PDF) inside the hadron and the parton to hadron fragmentation function (FF) etc.) are not calculated
at present from the first principle in QCD.

Because of this reason the non-perturbative PDF and FF are extracted from the experiments from the measurement of (physical) hadronic cross section by using the factorized formula \cite{ocs,ocs1,nst,pnr,pnl,pjet,pnhq,pncs,pncc,pngl,nrn}
\bea
d\sigma (H_1H_2 \rightarrow H_3+X) ~=~ \sum_{i,j,l}~\int dx_1 ~\int dx_2 ~\int dz~f_{i/H_1}(x_1,Q^2)~f_{j/H_2}(x_2,Q^2)~d{\hat \sigma}_{ij \rightarrow kl}~D_{H_3/k}(z,Q^2). \nonumber \\
\label{1o}
\eea
In eq. (\ref{1o}) the $f_{i/H_1}(x_1,Q^2)$ is the parton distribution function (PDF) of the parton $i$ inside the hadron $H_1$, the
$d{\hat \sigma}_{ij \rightarrow kl}$ is the short distance partonic level scattering differential cross section calculated by using pQCD, $D_{H_3/k}(z,Q^2)$ is the fragmentation function (FF) of the parton $k$ to fragment to the outgoing hadron $H_3$, the $Q$ is the factorization/renormalization scale, $x_1(x_2)$
is the longitudinal momentum fraction of the parton $i(j)$ with respect to the incoming hadron $H_1(H_2)$, $z$ is the longitudinal momentum fraction of the hadron $H_3$ with respect to the parton $k$, the $X$ represents the other (inclusive) outgoing hadrons and $i,j,k,l=q,{\bar q},g$ represent light quark, antiquark, gluon.

The non-perturbative parton distribution function (PDF) is an universal quantity, {\it i. e.}, it does not change from one experiment to another. Hence it is essential to derive the correct definition of the parton distribution function inside the hadron from the first principle in QCD. The correct definition of the parton distribution function inside the hadron must be consistent with the definition of the distribution function in quantum field theory because QCD is a quantum field theory.

Since the non-perturbative QCD is not solved yet, the operator product expansion (OPE) in QCD is assumed to be valid to study the QCD phenomenology for various processes at high energy colliders. For example the operator product expansion (OPE) in QCD is assumed to be valid to define the PDF which is then extracted from the experiments by using eq. (\ref{1o}). The definition of the parton distribution function (PDF) inside the hadron consistent with the operator product expansion (OPE) in QCD is derived in \cite{ocs}.

However, in this paper, we show that the gauge invariant definition of the non-perturbative gluon distribution function inside the hadron consistent with the operator product expansion (OPE) in QCD at high energy colliders is not consistent with the gauge invariant definition of the non-perturbative gluon distribution function in quantum field theory. Since the operator product expansion is applicable at short distance the OPE in QCD does not solve the long distance confinement problem involving hadron in QCD where the non-perturbative QCD is applicable.

The paper is organized as follows. In section II we discuss the definition of the gluon distribution function inside the hadron at high energy colliders using quantum field theory (QCD). In section III we show that the unrenormalized QCD and the renormalized QCD predict the same hadronic cross section at all orders in coupling constant in QCD. In section IV we discuss the operator product expansion and the total cross section in electron-positron annihilation to hadrons. In section V we focus on the operator product expansion in QCD and the confinement in QCD. In section VI we discuss the definition of the gluon distribution function inside the hadron at high energy colliders using the operator product expansion (OPE). In section VII we show that the operator product expansion in QCD is not consistent with quantum field theory for the gluon distribution function. Section VIII contains conclusions.

\section{Definition of the gluon distribution function inside hadron using quantum field theory (QCD) }

In the factorized formula in eq. (\ref{1o}), the single parton $i$ (a quark or an antiquark or a gluon) from the incoming hadron $H_1$ interacts with another single parton $j$ (a quark or an antiquark or a gluon) from the other incoming hadron $H_2$ in the short distance partonic level cross section ${\hat \sigma}_{ij \rightarrow kl}$ to produce the parton $k$ (a quark or an antiquark or a gluon) which fragments to the outgoing hadron $H_3$. Hence the parton distribution function (PDF) $f_{i/H_1}(x_1,Q^2)$ represents the probability of finding the single parton $i$ (a quark or an antiquark or a gluon) inside the incoming hadron $H_1$ in eq. (\ref{1o}). Similarly the parton to hadron fragmentation function (FF) $D_{H_3/k}(z,Q^2)$ represents the probability of a single parton $k$ (a quark or an antiquark or a gluon) fragmenting to the outgoing hadron $H_3$ in eq. (\ref{1o}).

In quantum mechanics the probability of finding a particle can be obtained from the quantum wave function of the particle. Hence in quantum mechanics the probability of finding a particle is proportional to $|\phi(x)|^2$ where $\phi(x)$ is the quantum wave function of the particle. Therefore in quantum field theory one expects that the  distribution function of the particle be proportional to the correlation function of the type $\phi^\dagger(x)\psi(0)$. In this section we will derive the definition of the quark distribution function inside the hadron and the definition of the gluon distribution function inside the hadron by using the quantum field theory.

In quantum field theory the free Dirac field containing positive and negative energy solution can be written as \cite{man}
\bea
&& \sum_{s=1}^2 \int \frac{d^3k}{(2\pi)^3\sqrt{2E_k}} [a_s(k) u_s(k) e^{-ik\cdot x} +b^\dagger_s(k) v_s(k) e^{ik\cdot x}]=\psi_q(x)+\psi^\dagger_{\bar q}(x)
\label{dwf}
\eea
where $a(k)$ is the annihilation operator of the quark and $b^\dagger(k)$ is the creation operator of the antiquark. The equal-time anti-commutation relations at the initial time, say at $t=t_{in}=0$ is given by
\bea
&&\{a_r(p), a^\dagger_s(p')\} = (2\pi)^3~ \delta_{rs}~\delta^{(3)}({\vec p}-{\vec p}'),~~~~~~~~~~~\{a_r(p), a_s(p')\} =0,~~~~~~~~~~~~\{a^\dagger_r(p), a^\dagger_s(p')\} =0\nonumber \\
&&\{b_r(p), b^\dagger_s(p')\} = (2\pi)^3~ \delta_{rs}~\delta^{(3)}({\vec p}-{\vec p}'),~~~~~~~~~~~~\{b_r(p), b_s(p')\} =0,~~~~~~~~~~~~~\{b^\dagger_r(p), b^\dagger_s(p')\} =0.\nonumber \\
\label{d3}
\eea
Dirac spinors satisfy
\bea
&&\sum_{s=1}^2 u_s(p) {\bar u}_s(p) = \pslash +m,~~~~~~~~~~~\sum_{s=1}^2 v_s(p){\bar v}_s(p) = \pslash -m \nonumber \\
&& u^\dagger_r(p) \cdot u_s(p) = v^\dagger_r(p) v_s(p) =\delta_{rs} 2E_p,~~~~~~~~~~~u^\dagger_r({\vec p})\cdot v_s(-{\vec p}) = v^\dagger_r({\vec p}) \cdot u_s(-{\vec p})=0.
\label{d4}
\eea
The number operator ${\hat n}(k)$ of a particle in quantum field theory is given by
\bea
{\hat n}(k)=a^\dagger(k)(k)
\label{nop}
\eea
and the distribution function $f(p)$ of the particle inside the hadron $H$ in quantum field theory is given by \cite{nc}
\bea
f(p)~ (2\pi)^3~ \delta^{(3)}({\vec p}-{\vec k}) = <H|a^\dagger(p) a(k)|H>
\label{ds1}
\eea
where $a^\dagger(k)$ is the creation operator of the particle and $a(k)$ is the annihilation operator of the particle.

Let us derive the following equation for the quark case at the initial time $t=t_{in}=0$  
\bea
&& \int d^3x e^{-i{\vec k} \cdot {\vec x}} <H|\psi^\dagger(x) \psi(0)|H> = \int d^3x \sum_{r=1}^2 \sum_{s=1}^2  \int \frac{d^3p}{(2\pi)^3 \sqrt{2E_p}} \int \frac{d^3p'}{(2\pi)^3 \sqrt{2E_{p'}}}<H|a^\dagger_r(p)a_s(p')|H> \nonumber \\
&&\times u^\dagger_r(p) \cdot u_s(p')e^{i({\vec p}-{\vec k}) \cdot {\vec x}}\nonumber \\
&&= \sum_{r=1}^2 \sum_{s=1}^2 \int \frac{d^3p'}{(2\pi)^3 \sqrt{4E_{p'}E_k}}<H|a^\dagger_r(k)a_s(p')|H> u^\dagger_r(k) \cdot u_s(p').
\label{d6}
\eea

From eq. (\ref{ds1}) we find for the quark case
\bea
f_{q/H}({\vec p})~ (2\pi)^3~ \delta_{rs}~\delta^{(3)}({\vec p}-{\vec p}') = <H|a^\dagger_r(p)a_s(p')|H>
\label{d7}
\eea
where $f_{q/H}({\vec p})$ is the quark distribution function inside the hadron $H$.

Using eqs. (\ref{d7}) and (\ref{d4}) in (\ref{d6}) we find
\bea
&&f_{q/H}({\vec p})= \frac{1}{2} \int d^3x e^{-i{\vec p} \cdot {\vec x}} <H|\psi^\dagger({\vec x}) \psi(0)|H>.
\label{d9}
\eea
In quantum field theory the eq. (\ref{d9}) for the free field theory can be extended to the interacting field theory (to the full QCD) by replacing the free fields by the interacting fields. Hence from eq. (\ref{d9}) we find that the gauge non-invariant definition of the quark distribution function $f_{q/H}( p)$ inside the hadron $H$ which is consistent with the definition of the distribution function of the quark in quantum field theory in full (interacting) QCD is given by
\bea
&&f_{q/H}(p)= \frac{1}{2} \int d^3x [e^{i p \cdot x} <H|\psi^{\dagger i}(x) \psi^i(0)|H>]_{t=0}.
\label{d10}
\eea
where $\psi^i(x)$ is in the full (interacting) QCD.

Similar to quark we find for the antiquark case at the initial time $t=t_{in}=0$ 
\bea
&& \int d^3x e^{-i{\vec k} \cdot {\vec x}} <H|\psi(x)\psi^\dagger(0) |H> \nonumber \\
&&= \sum_{r=1}^2 \sum_{s=1}^2 \int \frac{d^3p'}{(2\pi)^3 \sqrt{4E_{p'}E_k}}<H|b^\dagger_s(p')b_r(k)|H> v^\dagger_r(k) \cdot v_s(p').
\label{d6a}
\eea

Similar to eq. (\ref{d7}) we find for the antiquark case
\bea
&&f_{{\bar q}/H}({\vec p}) ~(2\pi)^3~ \delta_{rs}~ \delta^{(3)}({\vec p}-{\vec p}') = <H|b^\dagger_r({\vec p})b_s({\vec p}')|H>
\label{d7a}
\eea
where $f_{{\bar q}/H}({\vec p})$ is the antiquark distribution function inside the hadron $H$.

Using eqs. (\ref{d7a}) and (\ref{d4}) in (\ref{d6a}) we find
\bea
&&f_{{\bar q}/H}({\vec p})= \frac{1}{2} \int d^3x e^{-i{\vec p} \cdot {\vec x}} <H|\psi({\vec x}) \psi^{\dagger }(0)|H>.
\label{d9a}
\eea

Extending the free field theory equation (\ref{d9a}) to interacting field theory one finds that the gauge non-invariant definition of the antiquark distribution function $f_{{\bar q}/H}( p)$ inside the hadron $H$ which is consistent with the definition of the distribution function of the antiquark in quantum field theory in full (interacting) QCD is given by
\bea
f_{{\bar q}/H}( p)= \frac{1}{2} \int d^3x [e^{i p \cdot  x} <H|\psi^{i}(x) \psi^{\dagger i}(0)|H>]_{t=0}
\label{d10a}
\eea
where $\psi^{i}(x)$ is in the full (interacting) QCD.

In order to derive the definition of the gluon distribution function inside the hadron $H$ in QCD using quantum field theory let us consider the massless scalar gluon case before considering the gluon distribution function in QCD. Similar to the non-interacting quark and antiquark case discussed above one finds that the distribution function $f({\vec p})$ of massless scalar gluon inside the hadron in non-interacting quantum field theory is given by
\bea
&& f({\vec p})=2E_p \int d^3 x e^{-i{\vec p} \cdot {\vec x}} <H|\phi({\vec x}) \phi(0)|H>
\label{s6z}
\eea
where $\phi(x)$ is the massless scalar gluon field in the non-interacting quantum field theory.

Extending massless free scalar gluon field equation (\ref{s6z}) to massless interacting scalar gluon field we find that the definition of the scalar gluon distribution function $f( p)$ inside the hadron $H$ which is consistent with the definition of the distribution function of the scalar gluon in interacting quantum field theory is given by
\bea
&& f({\vec p})=2E_p \int d^3 x [e^{i p \cdot x} <H|\phi(x) \phi(0)|H>]_{t=0}
\label{s6}
\eea
where $\phi(x)$ is the massless scalar gluon field in the interacting quantum field theory.

Note that the massless scalar gluon field or the scalar gluon distribution function inside the hadron $H$ does not correspond to any physical situation. We have considered it here for simplicity to derive the gluon distribution function inside the hadron $H$ in QCD.

Hence extending eq. (\ref{s6}) for the scalar gluon distribution function to the gluon distribution function $f_{{ g}/H}(p)$ inside the hadron in QCD we find that the gauge non-invariant definition of the gluon distribution function $f_{{ g}/H}(p)$ inside the hadron $H$ in QCD obtained by using the quantum field theory is given by \cite{pjet}
\bea
f_{{ g}/H}( p)= 2E_p \int d^3x [e^{i p \cdot  x} <H|Q_\nu^c( x) Q^{\nu c}(0)|H>]_{t=0}
\label{g2}
\eea
where $Q^{\nu c}(x)$ is the (quantum) gluon field in the full (interacting) QCD with $\nu=0,1,2,3$ being the Lorentz index and $c=1,...,8$ being the color index.

Note that the definition of the quark, antiquark and gluon distribution functions in eqs. (\ref{d10}), (\ref{d10a}) and (\ref{g2}) are not gauge invariant. By using the path integral formulation of the background field method of QCD in the presence of SU(3) pure gauge background field $A_\mu^a(x)$ we have proved the factorization of soft and collinear divergences in QCD at all orders in coupling constant in \cite{pnr,pnl,pjet,pnhq,pncs,pncc}. Hence from eq. (\ref{d10}) we find that the gauge invariant definition of the quark distribution function inside hadron at high energy colliders which is consistent with the definition of the distribution function in quantum field theory and is consistent with the factorization of soft and collinear divergences in QCD at all orders in the coupling constant is given by \cite{pnl,nrn}
\bea
f_{q/H}(x) = \frac{1}{4\pi} \int dy' e^{-ixp^+y'} <H|{\bar \psi}(0,y',0_T)~\gamma^+~[{\cal P}e^{igT^{c}\int_0^{y'} dy'' A^{+c}(0,y'',0)}]~ \psi(0)|H> \nonumber \\
\label{o67}
\eea
where $A_\mu^a(x)$ is the SU(3) pure gauge background field.

Similarly from eq. (\ref{g2}) we find that the gauge invariant definition of the gluon distribution function inside hadron at high energy colliders
which is consistent with the definition of the distribution function in quantum field theory and is consistent with the factorization of soft and collinear divergences in QCD at all orders in the coupling constant is given by \cite{pjet,pngl,nrn}
\bea
f_{g/H}(x) = \frac{p^+}{2\pi} \int dy' e^{-ixp^+y'} <H|Q_\nu^b(0,y',0_T)~[{\cal P}e^{igT^{(A)c}\int_0^{y'} dy'' A^{+c}(0,y'',0)}]~ Q^{\nu b}(0)|H>. \nonumber \\
\label{o68}
\eea
Since the (quantum) gluon field $Q_\mu^a(x)$ in eq. (\ref{o68}) transforms gauge covariantly under the type I gauge transformation in the background field method of QCD \cite{oab} the definition of the gluon distribution function in eq. (\ref{o68}) is gauge invariant \cite{pjet,pngl}.

In eqs. (\ref{o67}) and (\ref{o68}) the Wilson line contains the SU(3) pure gauge background field $A_\mu^a(x)$. Since $A_\mu^a(x)$ is the SU(3) pure gauge background field it gives vanishing field tensor
\bea
F_{\lambda \mu}^a(x)=0,~~~~~~~~~~~~F_{\lambda \nu}^b(x) = \partial_\lambda A_\nu^b(x) - \partial_\nu A_\lambda^b(x) + gf^{bad} A_\lambda^a(x) A_\nu^d(x)
\eea
which means the SU(3) pure gauge background field $A_\mu^a(x)$ does not contribute to the physical cross section. The only role of the SU(3) pure gauge background field $A_\mu^a(x)$ in eqs. (\ref{o67}) and (\ref{o68}) is to maintain the gauge invariance of the quark and gluon distribution functions and to prove the factorization of soft and collinear divergences in QCD at all orders in coupling constant. The color field also plays an important role to study production of quark-gluon plasma in the laboratory \cite{png5,png6,png7,png8}.

It is important to note that since the SU(3) pure gauge background field $A_\mu^a(x)$ is the classical field one finds that the definition of the gauge invariant quark distribution function in eq. (\ref{o67}) contains quadratic powers of the (quantum) quark field $\psi(x)$ consistent with the gauge invariant definition of the distribution function of quark in quantum field theory even after the Wilson line is supplied in the gauge invariant definition of the quark distribution function in eq. (\ref{o67}).

Similarly since the SU(3) pure gauge background field $A_\mu^a(x)$ is the classical field one finds that the definition of the gauge invariant gluon distribution function in eq. (\ref{o68}) contains quadratic powers of the (quantum) gluon field $Q_\mu^a(x)$ consistent with the definition of the gauge invariant gluon distribution function in quantum field theory even after the Wilson line is supplied in the gauge invariant definition of the gluon distribution function in eq. (\ref{o68}).

Note that the quark distribution function in eq. (\ref{o67}) in the interacting (full) QCD is infinite in the unrenormalized QCD. In the renormalized QCD the renormalized quark distribution function is finite. Hence one may argue that since the quark distribution function in the unrenormalized QCD is divergent it is not necessary to define a quark distribution function in the unrenormalized QCD consistent with the definition of the distribution function of the quark in quantum field theory. However, this argument is not correct because in the next section we will show that the renormalized quark distribution function in the renormalized QCD and the unrenormalized quark distribution function in the unrenormalized QCD predict the same hadronic cross section at all orders in  coupling constant in QCD. This is consistent with the fact that the quark is not directly experimentally observed but the hadron is directly experimentally observed.

Similarly the gluon distribution function in eq. (\ref{o68}) in the interacting (full) QCD is infinite in the unrenormalized QCD. In the renormalized QCD the renormalized gluon distribution function is finite. Hence one may argue that since the gluon distribution function in the unrenormalized QCD is divergent it is not necessary to define a gluon distribution function in the unrenormalized QCD consistent with the definition of the distribution function of the gluon in quantum field theory. However, this argument is not correct because in the next section we will show that the renormalized gluon distribution function in the renormalized QCD and the unrenormalized gluon distribution function in the unrenormalized QCD predict the same hadronic cross section at all orders in  coupling constant in QCD. This is consistent with the fact that the gluon is not directly experimentally observed but the hadron is directly experimentally observed.

\section{ Unrenormalized QCD and Renormalized QCD Predict Same Hadronic Cross Section at All Orders in Coupling Constant}\label{rnurn}

In eqs. (\ref{o67}) and (\ref{o68}) we have derived the gauge invariant definition of the quark and gluon distribution functions inside the hadron which are consistent with the definition of the distribution functions of the quark and gluon in the quantum field theory but these distribution functions in eqs. (\ref{o67}) and (\ref{o68}) are divergent in the unrenormalized QCD. Hence one may argue that since the parton distribution function in the unrenormalized QCD is divergent it is not necessary to define a gluon distribution function in the unrenormalized QCD consistent with the definition of the distribution function of the gluon in quantum field theory. However, this argument is not correct because we will show in this section that the definition of the gluon distribution function in the unrenormalized QCD can be used to predict the correct hadronic cross section at all orders in coupling constant in QCD (see eq. (\ref{1orn})).

Using the path integral formulation of the background field method of QCD in the presence of SU(3) pure gauge background field we have simultaneously proved the renormalization of ultra violet (UV) divergences and the factorization of infrared (IR) and collinear divergences in QCD at all orders in coupling constant in \cite{nrn}. Hence from eq. (\ref{1o}) and \cite{nrn} we find
\bea
&&d\sigma (H_1H_2 \rightarrow H_3+X) ~=~ \sum_{i,j,l}~\int dx_1 ~\int dx_2 ~\int dz~f_{i/H_1}(x_1,Q^2)~f_{j/H_2}(x_2,Q^2)~d{\hat \sigma}_{ij \rightarrow kl}~D_{H_3/k}(z,Q^2) \nonumber \\
&&=~ \sum_{i,j,l}~\int dx_1 ~\int dx_2 ~\int dz~f^{\rm UnRenormalized}_{i/H_1}(x_1,Q^2)~f^{\rm UnRenormalized}_{j/H_2}(x_2,Q^2)~d{\hat \sigma}^{\rm UnRenormalized}_{ij \rightarrow kl}\nonumber \\
&&\times D^{\rm UnRenormalized}_{H_3/k}(z,Q^2)
\label{1orn}
\eea
where $f(x,Q^2)$ is the renormalized parton distribution function in the renormalized QCD, $f^{\rm UnRenormalized}(x,Q^2)$ is the unrenormalized parton distribution function in the unrenormalized QCD, $D(z,Q^2)$ is the renormalized fragmentation function in renormalized QCD, $D^{\rm UnRenormalized}(z,Q^2)$ is the unrenormalized fragmentation function in unrenormalized QCD, ${\hat \sigma}$ is the renormalized partonic level cross section in renormalized QCD and ${\hat \sigma}^{\rm UnRenormalized}$ is the unrenormalized partonic level cross section in the unrenormalized QCD.

Hence one finds from eq. (\ref{1orn}) that the correct definition of the parton distribution function consistent with the definition of the distribution function from the first principle in the quantum field theory plays a very important role to prove that the hadronic cross section in the renormalized QCD is exactly same as the hadronic cross section in the unrenormalized QCD at all orders in coupling constant. Note that we are not saying that one does not have to do renormalization in QCD at fixed orders of coupling constant calculation but what we are saying is that at all orders of coupling constant the renormalized QCD and the unrenormalized QCD predict the same hadronic cross section. It is human limitations that we can not perform all orders coupling constant calculation in QCD but nature does not work according to human limitations. The QCD as a fundamental theory of the nature predicts that the renormalized QCD and the unrenormalized QCD predict the same hadronic cross section at all orders in coupling constant. This is consistent with the fact that the quarks and gluons are not directly experimentally observed but the hadrons are directly experimentally observed.

\section{ Operator Product Expansion and Total Cross Section in Electron-Positron Annihilation to Hadrons }

Since there are no fragmentation functions in the total cross section in electron-positron annihilation to hadrons one may argue that the eq. (\ref{1orn}) is not applicable to study the total cross section in electron-positron annihilation to hadrons. Hence one may argue that the renormalized QCD and the unrenormalized QCD predict the same hadronic cross section at all orders in coupling constant is wrong. However, this argument is not correct which can be seen as follows.

Since one calculates the total cross section in electron-positron annihilation to hadrons one uses in the renormalized QCD
\bea
\sum_H {D_i^H} =1
\label{drenh}
\eea
where $D_i^H$ is the renormalized fragmentation function for the parton $i$ to fragment to hadron $H$. Using eq. (60) of \cite{nrn} in (\ref{drenh}) we find in the unrenormalized QCD that
\bea
\sum_H {D_i^H}^{\rm UnRenormalized}=Z^2
\label{durnh}
\eea
where $Z$ is the (quantum) field divergent renormalization factor of the parton $i$. Hence the divergent renormalization factor $Z$ in eq. (\ref{durnh}) exactly cancels with the corresponding divergent factor $Z$ in the partonic level cross section $\sigma_{e^+e^-\rightarrow k_1,k_2,...,k_n}$ in the electron-positron annihilation to partons at all orders of coupling constant in the unrenormalized QCD, similar to eq. (\ref{1orn}).

Hence one finds that the total cross section in the electron-positron annihilation to hadrons is finite at all orders in coupling constant in the unrenormalized QCD and is exactly the same total cross section that is obtained in the electron-positron annihilation to hadrons in the renormalized QCD at all orders in coupling constant.

Because of eq. (\ref{drenh}) there are no fragmentation functions appearing in the total cross section in the inclusive electron-positron annihilation to hadrons in the renormalized QCD. Hence the current-current correlator that appears in the operator product expansion (OPE) to study the total cross section in the electron-positron annihilation to hadrons becomes the vacuum expectation of the form
\bea
<0|J_\mu(z) j_\nu(y)|0>= C(z-y) <0|O_{\mu ... \nu}(y)|0>
\label{opep}
\eea
where $C(z-y)$ is the short distance coefficient which is singular as the distance $(z-y)^\mu \rightarrow 0$ and $O_{\mu ... \nu}(y)$ is the local operator.

Since the vacuum expectation is used in eq. (\ref{opep}) instead of the hadronic expectation $<H|O_{\mu ... \nu}(y)|H>$ one finds that the long distance confinement involving hadron in QCD does not play any role to study the total cross section in the electron-positron annihilation to hadrons. This implies that the use of the operator product expansion at short distance \cite{okw} is ok to study the total cross section in the electron-positron annihilation to hadrons for high momentum transfer processes.

\section{Operator Product Expansion and Confinement in QCD Involving Hadron }

For the hadronic expectation $<H|O_{\mu ... \nu}(y)|H>$ the long distance confinement involving hadron $H$ in QCD must be included. Since the operator product expansion (OPE) is applicable at short distance the OPE in QCD does not solve the long distance confinement problem involving hadron in QCD where the non-perturbative QCD is applicable

The inclusive cross section with identified hadron $H$ in the electron-positron annihilation involves the hadronic expectation of the current-current correlator of the form
\bea
\sum_X <0|J_\mu(y)|H+X><H+X| j_\nu(0)|0>=<0|J_\mu(y)a^\dagger_H a_H j_\nu(0)|0>,~~~~~~~~~~~|H+X>=a^\dagger_H |X>\nonumber \\
\label{opep1}
\eea
where $a^\dagger_H$ is the creation operator of the hadron. In eq. (\ref{opep1}) the operator product expansion (OPE) does not apply and no short distance analysis exists for the case of inclusive $e^+e^- \rightarrow H+X$ process \cite{jaftw}.

The first principle issue here is that the operator product expansion (OPE) in QCD is valid at short distance where pQCD is applicable but the confinement in QCD involving hadron occurs at long distance where the non-perturbative QCD is applicable. Hence the OPE is applied to short distance pQCD calculation which does not require any information about long distance non-perturbative QCD confinement involving hadron, such as the total cross section in the electron-positron annihilation to hadrons \cite{jaftw}. However, any calculation in pQCD to study physical phenomena which requires long distance non-perturbative QCD confinement involving hadron may not always be correctly studied from the first principle in QCD by the operator product expansion (OPE) alone without incorporating the non-perturbative QCD which is not solved yet.

\section{Definition of the gluon distribution function inside hadron using operator product expansion }

Let us now discuss the definition of the gluon distribution function consistent with the operator product expansion (OPE) in QCD which is widely used in the literature at high energy colliders. The standard procedure in the operator product expansion in the deep inelastic scattering involving lepton and hadron is as follows: The structure functions $F_1$ and $F_2$ are related to the product of electromagnetic currents inside the hadron as \cite{fw}
\bea
\frac{1}{2\pi} \int d^4y e^{iq \cdot y} <H|j_\mu(y) j_\lambda(0) |H> =-\frac{g_{\mu \lambda}}{m} F_1(\nu,q^2)+\frac{p_\mu p_\lambda}{m\nu } F_2(\nu, q^2),~~~~~~~\nu=p\cdot q
\eea
where $p^\mu$ is the hadron momentum and $q^\mu$ is the momentum transfer to the hadrons. The product of two currents at short distances in the OPE is given by \cite{fw}
\bea
j_\mu(y) j_\lambda(0) \propto C(y)~O_{\mu ... \lambda}(0)
\eea
where $C(y)$ is the short distance coefficient and $O_{\mu ... \lambda}$ is the local operator.

Let us consider the gluon sector. Since the quantum gluon field $Q_\mu^a(y)$ is not gauge covariant in QCD the twist two operator that appears in the operator product expansion of two currents is given by \cite{fw,ocs}
\bea
{\cal O}_g^{\mu...\nu}(0)\equiv \frac{1}{2} {\rm Tr}[F^{\mu \delta}(0) iD^\lambda[Q](0)...iD^\beta[Q](0) F_{\delta}^\nu(0)],~~~~~~~~D_\mu^{ab}[Q] = \delta^{ab} \partial_\mu + gf^{acb} Q_\mu^c\nonumber \\
\label{o9g}
\eea
which contains gluon field tensor $F_{\mu \nu}^a(y)$ instead of gluon field $Q_\mu^a(y)$ where
\bea
F_{\mu \lambda}^a(y) = \partial_\mu Q_\lambda^a(y) - \partial_\lambda Q_\mu^a(y) + gf^{acd} Q_\mu^c(y) Q_\lambda^d(y)
\label{g4}
\eea
which transforms gauge covariantly in QCD.

In particular the hadronic matrix element of the gauge invariant twist two operator
\bea
<H|{\cal O}_g^{\mu...\nu}(0)|H>\equiv \frac{1}{2} <H|{\rm Tr}[F^{\mu \delta}(0) F_{\delta}^\nu(0)]|H>
\label{o9ga}
\eea
corresponds to the moment of the gauge invariant gluon distribution function $f_{g/H}(x)$ inside the hadron \cite{ocs}
\bea
\int_0^1 dx f_{g/H}(x) \equiv <H|{\cal O}_g^{\mu...\nu}(0)|H>\equiv \frac{1}{2} <H|{\rm Tr}[F^{\mu \delta}(0) F_{\delta}^\nu(0)]|H>
\eea
if the gauge invariant definition of the gluon distribution function $f_{g/H}(x)$ inside the hadron is given by \cite{ocs}
\bea
f_{g/H}(x) = \frac{1}{2\pi xp^+} \int dy' e^{-ixp^+y'} <H|F^{+\nu b}(0,y',0_T)~[{\cal P}e^{igT^{(A)c}\int_0^{y'} dy'' Q^{+c}(0,y'',0)}]~ F_\nu^{+ b}(0)|H>. \nonumber \\
\label{o69}
\eea
Note that the Wilson line in eq. (\ref{o69}) contains quantum gluon field $Q_\mu^a(y)$ whereas the Wilson line in eq. (\ref{o68}) contains classical SU(3) pure gauge background field $A_\mu^a(y)$.

It is useful to mention that one can not use the classical SU(3) pure gauge background field $A_\mu^a(y)$ in the Wilson line in eq. (\ref{o69}) because then $f_{g/H}(x)$ in eq. (\ref{o69}) will not be gauge invariant and will not be consistent with the factorization of soft and collinear divergences at all orders in the coupling constant in QCD. This is because in the background field method of QCD in the presence of SU(3) pure gauge background field $A_\mu^a(y)$ the gauge invariant gluon field is given by \cite{pjet,pngl,nrn}
\bea
[{\cal P}e^{-igT^{(A)c}\int_{y'}^\infty dy'' A^{+c}(0,y'',0)}]Q_\mu^b(0,y',0_T)
\eea
where the soft and collinear divergences are factorized into the exponential containing the SU(3) pure gauge background field $A_\mu^a(y)$.

Hence we find that the gauge invariant definition of the gluon distribution function inside the hadron consistent with the operator product expansion (OPE) in QCD is given by eq. (\ref{o69}).

\section{ Operator Product Expansion in QCD Is Not Consistent With Quantum Field Theory For Gluon Distribution Function }

From eq. (\ref{o69}) one finds that there is no way the gauge invariant definition of the gluon distribution function $f_{g/H}(x)$ in eq. (\ref{o69}) can be called as a distribution function in quantum field theory because it contains infinite powers of quantum gluon field $Q_\mu^a(x)$ instead of quadratic powers of the quantum gluon field $Q_\mu^a(x)$. Even without the Wilson line the gauge non-invariant definition from eq. (\ref{o69}) contains cubic and quartic powers of the quantum gluon field $Q_\mu^a(x)$ and hence can not be called as a distribution function in quantum field theory.

One can note that the definition of $f_{g/H}(x)$ in eq. (\ref{o69}) corresponds to distribution function in quantum field theory in the light-cone gauge $Q^+=0$ because in the light-cone gauge it contains the quadratic powers of the quantum gluon field $Q_\mu^a(x)$. However, in any other gauge the gauge invariant definition of $f_{g/H}(x)$ in eq. (\ref{o69}) does not correspond to distribution function in quantum field theory because in any other gauge (except the light-cone gauge) it contains infinite powers of quantum gluon field $Q_\mu^a(x)$ instead of quadratic powers of the quantum gluon field $Q_\mu^a(x)$. As mentioned above even without the Wilson line the gauge non-invariant definition from eq. (\ref{o69}) contains cubic and quartic powers of the quantum gluon field $Q_\mu^a(x)$ and hence can not be called as a distribution function in quantum field theory.

Since the definition of the gluon distribution function $f_{g/H}(x)$ in eq. (\ref{o69}) is gauge invariant it must correspond to the definition of the distribution function in any gauge. Hence we find that the gauge invariant definition of the gluon distribution function in eq. (\ref{o69}) which is consistent with the operator product expansion (OPE) in QCD is not consistent with definition of the gauge invariant gluon distribution function in quantum field theory.

On the other hand the gauge invariant definition of the gluon distribution function in eq. (\ref{o68}) contains quadratic powers of the quantum gluon field $Q_\mu^a(x)$ [even after the Wilson line is supplied] which implies that the gauge invariant definition of the gluon distribution function in eq. (\ref{o68}) is consistent with the definition of the gauge invariant gluon distribution function in quantum field theory.

As mentioned in section \ref{rnurn} one may argue that since the gluon distribution function in the unrenormalized QCD is divergent it is not necessary to define a gluon distribution function in the unrenormalized QCD consistent with the definition of the distribution function of the gluon in quantum field theory. However, this argument is not correct because we have shown in section \ref{rnurn} that the definition of the gluon distribution function in the unrenormalized QCD can be used to predict the correct hadronic cross section at all orders in coupling constant in QCD, see eq. (\ref{1orn}) and  \cite{nrn}.

In summary we find that the gauge invariant definition of the non-perturbative gluon distribution function inside the hadron which is consistent with the operator product expansion (OPE) in QCD is not consistent with the gauge invariant definition of the non-perturbative gluon distribution function in the quantum field theory.

\section{Conclusions}
Since the operator product expansion (OPE) is applicable at short distance the OPE in QCD does not solve the long distance confinement problem involving hadron in QCD where the non-perturbative QCD is applicable. In this paper we have shown that the gauge invariant definition of the non-perturbative gluon distribution function inside the hadron consistent with the operator product expansion in QCD at high energy colliders is not consistent with the gauge invariant definition of the non-perturbative gluon distribution function in the quantum field theory.

\end{document}